\documentclass{elsart}

\usepackage{epsfig}

\usepackage{amssymb}

\begin{document}
\include{elsart-num.bst}
\begin{frontmatter}



\title{Weak Lensing from Space I:
       Instrumentation and Survey Strategy}


\author[cit,gsfc]{Jason Rhodes\thanksref{nrc}}
\address[cit]{California Institute of Technology, 1201 E. California
Blvd., Pasadena,          CA 91125}
\address[gsfc]{Laboratory for
Astronomy \& Solar Physics, Code 681,Goddard Space Flight Center,
Greenbelt MD 20771} \ead{rhodes@astro.caltech.edu}
\thanks[nrc]{NASA/NRC Research Associate}

\author[cit,sac]{Alexandre Refregier}
\address[sac]{Service d'Astrophysique, B\^{a}t. 709, CEA Saclay, F-91191
Gif sur Yvette, France}

\author[cam] {Richard Massey}
\address[cam]{Institute of Astronomy, Madingley Road, Cambridge CB3
OHA, U.K.}
\author[cit]{Justin Albert}
\author[5]{David Bacon}
\author[6]{Gary Bernstein}
\author[cit]{Richard Ellis}
\author[6]{Bhuvnesh Jain}
\author[7]{Alex Kim}
\author[8]{Mike Lampton}
\author[9]{Tim McKay}
\author[9]{C.~Akerlof}
\author[7]{G.~Aldering}
\author[10]{R.~Amanullah}
\author[11]{P.~Astier}
\author[yal]{C.~Baltay}
\author[11]{E.~Barrelet}
\author[7]{C.~Bebek}
\author[10]{L.~Bergstr\"{o}m}
\author[7]{J.~Bercovitz}
\author[8]{M.~Bester}
\author[9]{B.~Bigelow}
\author[15]{R.~Bohlin}
\author[12]{A.~Bonissent}
\author[13]{C.~Bower}
\author[9]{M.~Brown}
\author[9]{M.~Campbell}
\author[7]{W.~Carithers}
\author[8]{E.~Commins}
\author[7]{C.~Day}
\author[14]{S.~Deustua}
\author[7]{R.~DiGennaro}
\author[12]{A.~Ealet}
\author[yal]{W.~Emmet}
\author[10]{M.~Eriksson}
\author[17]{D.~Fouchez}
\author[15]{A.~Fruchter}
\author[11]{J-F.~Genat}
\author[9]{D.~Gerdes}
\author[6]{L.~Gladney}
\author[8]{G.~Goldhaber}
\author[10]{A.~Goobar}
\author[7]{D.~Groom}
\author[8]{ S.~Harris}
\author[8]{P.~Harvey}
\author[8]{H.~Heetderks}
\author[7]{S.~Holland}
\author[16]{D.~Huterer}
\author[7]{W.~Johnston}
\author[7]{A.~Karcher}
\author[7]{W.~Kolbe}
\author[7]{B.~Krieger}
\author[7]{G.~Kushner}
\author[7]{N.~Kuznetsova}
\author[7]{R.~Lafever}
\author[7]{J.~Lamoureux}
\author[7]{ M.~Levi}
\author[7]{E.~Linder}
\author[7]{S.~Loken}
\author[9]{W.~Lorenzon}
\author[17]{R.~Malina}
\author[17]{A.~Mazure}
\author[9]{S.~McKee}
\author[7]{R.~Miquel}
\author[yal]{N.~Morgan}
\author[10]{E.~M\"{o}rtsell}
\author[13]{N.~Mostek}
\author[13]{S.~Mufson}
\author[13]{J.~Musser}
\author[7]{P.~Nugent}
\author[7]{H.~Oluseyi}
\author[11]{R.~Pain}
\author[7]{N.~Palaio}
\author[8]{D.~Pankow}
\author[7]{S.~Perlmutter}
\author[8]{R.~Pratt}
\author[17]{E.~Prieto}
\author[yal]{D.~Rabinowitz}
\author[7]{K.~Robinson}
\author[7]{N.~Roe}
\author[6]{D.~Rusin}
\author[9]{M.~Schubnell}
\author[8]{M.~Sholl}
\author[18]{G.~Smadja}
\author[cit]{R.~Smith}
\author[8]{G.~Smoot}
\author[yal]{J.~Snyder}
\author[7]{A.~Spadafora}
\author[yal]{A.~Szymkowiak}
\author[9]{G.~Tarl\'e}
\author[cit]{K.~Taylor}
\author[17]{A.~Tilquin}
\author[9]{A.~Tomasch}
\author[7]{H.~von der Lippe}
\author[11]{D.~Vincent}
\author[7]{J-P.~Walder}
\author[7]{G.~Wang}

\address[5]{Institute for Astronomy, Blackford Hill,
Edinburgh EH9 3HJ, U.K.} \address[6]{ Department of Physics \&
Astronomy, University of Pennsylvania, 209 S.33$^{\rm rd}$ Street,
Philadelphia, PA 19104 } \address[7]{ Lawrence Berkeley National
Laboratory, 1 Cyclotron Road, Berkeley, CA 94720}
\address[8]{ Space Sciences Lab., University of California,
Berkeley CA 94720} \address[9]{Department of Physics, University
of Michigan, Ann Arbor,  MI 48109, USA} \address[10]{ University
of Stockholm, Stockholm, Sweden} \address[11]{ CNRS/IN2P3/LPNHE,
Paris, France} \address[12]{ CNRS/IN2P3/CPPM, Marseille, France}
\address[13]{ Indiana University, Bloomington IN, USA}
\address[14]{ American Astronomical Society, Washington DC, USA}
\address[15]{ Space Telescope Science Institute, Baltimore MD,
USA} \address[16]{ Case Western Reserve University, Cleveland OH,
USA} \address[17]{ CNRS/INSU/LAM, Marseille, France}
\address[18]{ CNRS/IN2P3/IPNL, Lyon, France} \address[yal]{Yale
University}

\begin{abstract}

A wide field space--based imaging telescope is necessary to fully
exploit the technique of observing dark matter via weak
gravitational lensing.  This first paper in a three part series
outlines the survey strategies and  relevant instrumental
parameters for such a mission. As a concrete example of hardware
design, we consider the proposed {\it Supernova/Acceleration
Probe} (SNAP). Using SNAP engineering models, we quantify the
major contributions to this telescope's Point Spread Function
(PSF). These PSF contributions are relevant to any similar wide
field space telescope. We further show that the PSF of SNAP or a
similar telescope will be smaller than current ground-based PSFs,
and more isotropic and stable over time than the PSF of the {\it
Hubble Space Telescope}. We outline survey strategies for two
different regimes -- a ``wide'' 300 square degree survey and a
``deep'' 15 square degree survey that will accomplish various weak
lensing goals including statistical studies and dark matter
mapping.

\end{abstract}
\begin{keyword}
dark matter \sep dark energy \sep instrumentation
\PACS 95.35.+d \sep 95.55.Fw \sep 95.55.-n \sep 98.80.-k
\end{keyword}
\end{frontmatter}


\section{Introduction} \label{introduction} A major thrust in cosmology is the
understanding of the dual phenomena of dark matter and dark
energy.  Over 60 years of increasingly convincing observations
have shown that most of the matter ($\sim90$\%) in the universe is
some form of non-baryonic dark matter \cite{tur00}.  The nature of
this dark matter and its relation to the baryonic matter
comprising stars and galaxies remain as crucial questions in
modern cosmology. More recently, several groups have used
observations of type Ia supernovae to demonstrate that the
expansion of the universe is accelerating \cite{rie98}
\cite{per99}. This surprising result points to the existence of a
dark energy with negative pressure driving the expansion of the
universe. These results are consistent with the `concordance
model' of a flat universe with critical density, consisting of
$\Omega_m\approx0.3$ and $\Omega_\Lambda\approx0.7$ \cite{pee02}.
It is clear that this `standard' universe is dominated by its
unknown dark components which are still not understood.

The importance of understanding the dark components of the
universe was stressed in a recent report made by the National
Research Council's Committee on the Physics of the Universe, which
listed dark matter and dark energy as two of the top
questions facing cosmology in the new millennium \cite{tur02}.
This committee recommended building a wide-field telescope in
space as a way to explore the dark energy.  An example of such a
telescope is the proposed {\it Supernova/Acceleration Probe}
(SNAP). The primary goal of SNAP is to study the accelerating
expansion of the universe and the nature of the dark energy, using
the same method by which the acceleration was discovered: type Ia
supernovae.

This is the first in a series of papers in which we demonstrate
that space-based observations by a wide-field telescope are useful
for studying the dark matter via weak gravitational lensing (see
also paper II \cite{mas03a} and paper III  \cite{ref03}). The
measurement of small distortions of the shapes of background
galaxies by foreground dark matter is an ideal method for
constraining the amount and distribution of dark matter in the
universe (e.g. see \cite{mel02} for a  review).

In this paper we study the hardware requirements of a wide field
space telescope for weak lensing.  As a concrete example, we use
the baseline telescope, optics, and filter specifications for
SNAP. We show that this hardware will achieve excellent image
quality over a wide field of view, with a low level of relevant
systematic effects compared to those in the \emph{Hubble Space
Telescope} (HST) or ground--based observatories. In
\S\ref{justification} we outline the reasons why weak lensing can
be measured so much more accurately from space. \S\ref{mission}
introduces the SNAP mission and hardware. \S\ref{strategy} covers
survey strategies. In \S\ref{systematic} we discuss the impact of
instrumental systematics including the point spread function
(PSF). Our conclusions are summarized in  \S\ref{conclusions}.

In paper II of this series, we  use detailed image simulations to
compute the efficiency  of weak lensing from space and we study
the prospects for making high resolution maps of the dark matter
distribution. In paper III, we  show the exquisite constraints
that a telescope such as SNAP can set on cosmological parameters
including $\Omega_m$ and the dark energy equation of state
parameter $w$.

\section{Why Space?}
\label{justification}

As we shall demonstrate in this series of papers, a wide field
space telescope is ideally suited to perform weak lensing studies.
Weak lensing measurements from the ground are fundamentally
limited by the relatively large and variable PSF introduced by
atmospheric seeing.  These limitations can be avoided in space but
 current space-based measurements are  limited by the small field of
view of \emph{HST}.

Future space missions like   SNAP are being designed from the
start to produce repeatable observations with excellent photometry
and imaging characteristics across a wide field of view.  Such
missions will provide precise  measurements of weak lensing shear
in galaxy shapes. With the precision photometry available from
space, it may also be possible to consider the lensing
magnification of background galaxies \cite{jai02}.

In the following, we use the specifications for SNAP as a concrete
example of a wide field space-based  telescope.  Any similar
telescope will have similar advantages over \emph{HST} and
ground-based telescopes.  It will  face similar engineering
requirements and technical constraints and be subject to analogous
systematic effects. Therefore, the results presented here are
specific to SNAP but remain directly
relevant to any generic wide field imager from space.

Compared to HST, SNAP has a wide field of view and high instrument
throughput, enabling it to efficiently survey the large area
needed to constrain cosmological parameters.  Due to its long
three day orbit and the facts that SNAP will rarely enter
Earth-shadow and will maintain one side facing the sun, SNAP will
also have greater thermal stability than HST. This leads to a more
constant and therefore better understood PSF. Hence, deconvolution
can be performed more accurately, and object shapes can be
corrected for the effects of PSF distortion with a lower level of
systematic errors. Relative to current and planned ground-based
observatories with wide fields of view, SNAP has a small PSF. This
leads to lower systematics even before correction, and to a higher
surface density of resolved galaxies. Because the average galaxy
size decreases with increasing redshift, SNAP is also able to
probe more distant galaxies than is possible from the ground. The
particular strengths of SNAP for weak lensing studies are thus:

\begin{itemize}
\item high surface density of resolved galaxies \item low
systematics due to small PSF and thermal stability \item extensive
filter set for calculation of photometric redshifts \item high
median redshift of resolved galaxies.
\end{itemize}

The strengths of SNAP outlined in the previous paragraph, and
expanded upon in \S3.1 of paper II, will provide SNAP with the
unique ability to address a variety of new science goals via weak
lensing.  These  goals include:

\begin{itemize}
\item creation of high resolution dark matter maps \item high
precision measurement of weak lensing statistics \item creation of
an extensive mass selected halo catalog \item precision
measurement of cosmological parameters including $\Omega_{M}$,
$\Omega_{\Lambda}$, $\sigma_{8}$, and the
      dark energy equation of state parameter $w$
\item measurement of the evolution of structure through 3-D
mapping and through
      the redshift dependence of lensing statistics
\item testing of the gravitational instability paradigm of
      structure formation.
\end{itemize}

As we demonstrate in this series of papers, these significant
goals can be accomplished only with the use of a space-based
wide-field observatory.

\section{The SNAP Mission} \label{mission}

SNAP is currently being designed for an approximately 40 month
mission. After an initial cool-down and calibration period, the
primary mission will be two deep 16 month supernova search
campaigns (one towards the northern hemisphere and one towards the
south) interspersed with a 5 month wide-field weak lensing survey.
After the 40 month design mission  SNAP may be operated as a guest
observer observatory on a competitive basis. For further details
of the SNAP mission see \cite{alc03}, \cite{ald02}, \cite{kim02},
\cite{lam02a}, \cite{lam02b}, \cite{tar02} and \cite{per02}.

The SNAP focal plane is partially covered by detectors using 6
optical filters spanning 350-1000~nm and 3 near infrared (NIR)
filters spanning 0.9-1.7~$\mu$m. SNAP will have 0.7 square degrees
of imaging coverage per pointing, half of it covered by optical
detectors and half by NIR detectors. The optical CCDs are being
designed at Lawrence Berkeley National Laboratory and the NIR
HgCdTe detectors will be like those used on the \emph{HST's} Wide
Field Camera 3.
All of the filters will be fixed in the focal plane, possibly by
attaching them permanently to the detectors.

\begin{figure}
 \includegraphics[width=15cm]{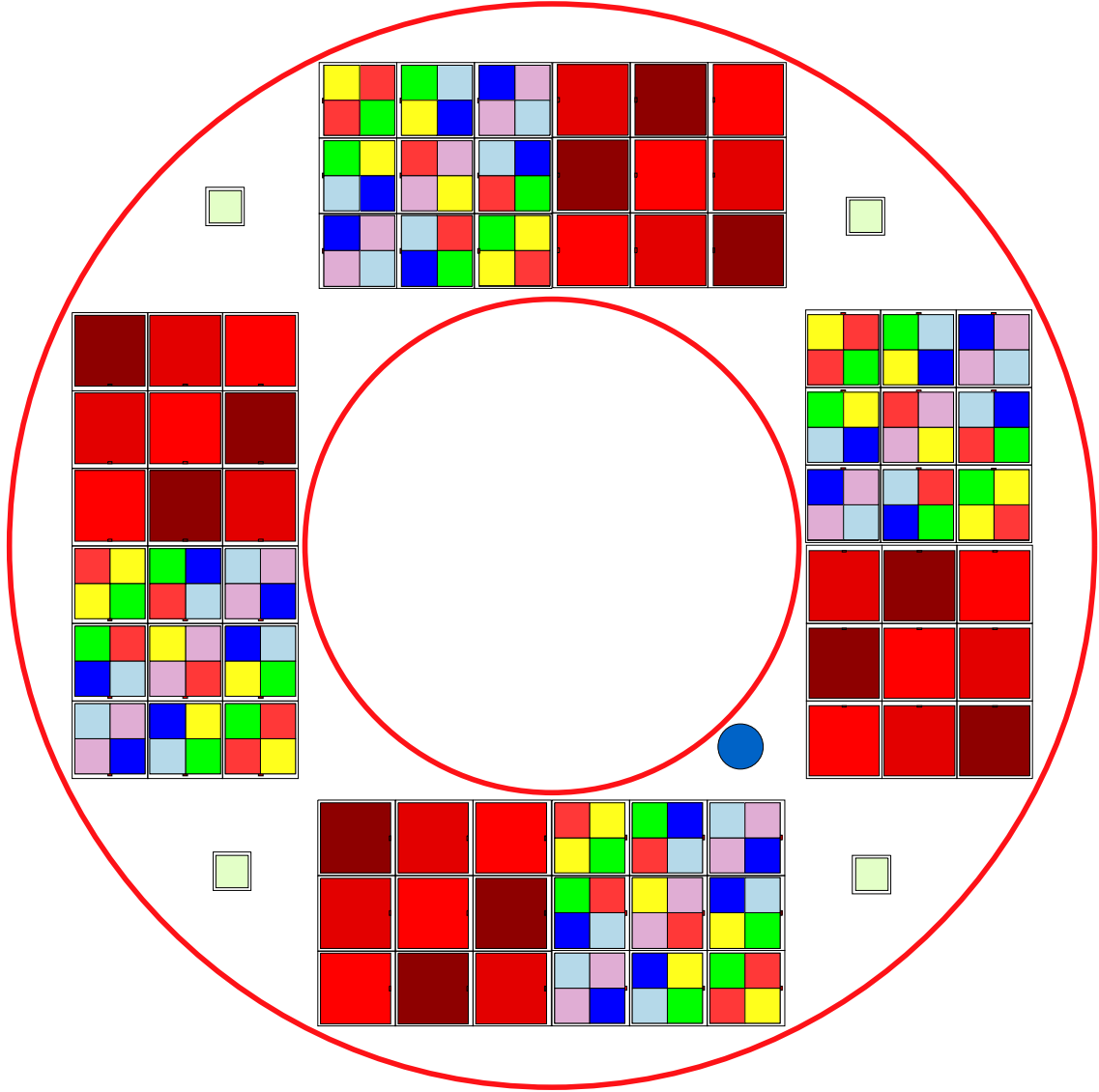}
 \caption{The layout of detectors on the SNAP focal plane.
 Each $17.5\times17.5$ square arcminute  CCD bank contains a $6\times6$ array
 of optical filters. Each infrared HgCdTe bank contains a $3\times3$ array with
 the same area. The total area of the detectors is 0.7 square degrees. The
 inner annulus has a radius of 0.06 radians ($0.34^{\circ}$) and the outer
 annulus has a radius of 0.013 radians ($0.74^{\circ}$). The spectrograph
 optical port is the small circle in the lower right quadrant. The four small
 squares are star guiders.}
 \label{fig:snap}
\end{figure}

The SNAP CCDs and HgCdTe detectors are arranged in an annulus in
the SNAP focal plane. As shown in Figure~\ref{fig:snap}, there are
four banks of CCDs and four banks of HgCdTe detectors.  Each bank
of CCDs consists of an array of $3\times3$ CCDs.  Each CCD is then
covered by a $2\times2$ grid of optical filters, in quarters of
different colors.  Thus, each CCD bank is a $6\times6$ array of
optical filters.  The pattern of colors is arranged so that as the
telescope is slewed across the sky either horizontally or
vertically, each patch of sky will be viewed through all 6 optical
filters in turn. A step-and-stare technique, whereby the telescope
is slewed repeatedly by the angular size of one optical filter
($\sim 3'$), accumulates an image in all bands without recourse to
a moving filter wheel.

Adjacent on the focal plane, the HgCdTe detectors are in
$3\times3$ arrays, with each detector covered entirely by just one
NIR filter (\emph{H', J }or \emph{K}).  Conveniently, since each
of the NIR filters is four times the area of a single optical
filter, the stacked NIR exposures in each sweep are twice as long
as the optical exposures.  As before, these filters are also
arranged so that one sweep will observe the same survey area in
all the filters, save for edge effects (the first and last fields
in the sweep direction will not be observed in both the infrared
and optical bands).


\section{Survey Strategy} \label{strategy}

\subsection{Deep Survey} \label{deep}

Approximately 60\% of the observing time in the two 16 month
supernova campaigns will be spent on photometry. A total of 15
square degrees (7.5 square degrees in each campaign) will be
scanned once every four days, stepping through all of the nine
filters.  Over the course of the deep survey, the total
integration time will be 144,000 seconds in each optical filter
and twice that in each infrared filter.  The remaining 40\% of the
time will be spent using the spectrograph to observe approximately
1000 supernovae per field (2000 total supernovae) that will be
detected out to $z\approx1.7$. During spectroscopy, the imagers
will be left switched on and any coincidental further integration
within the survey region will be in addition to these numbers.

The deep survey will be useful for several weak lensing studies.
The extremely high number density of resolved background galaxies
($\sim 260$ per square arcminute), each with a local shear
estimator, samples the lensing field with very high resolution. As
described in paper II, this can be converted into a detailed
two-dimensional (projected) map of the mass distribution which
shows clusters, filaments, and structure down to the scale of
galaxy groups. The nine filters will provide photometric redshifts
for almost all these galaxies, accurate to $\Delta z\approx 0.02$
(paper II). This will allow the subdivision of the detected
galaxies into redshift bins in order to trace the evolution of the
mass power spectrum. Furthermore, recent theoretical developments
make possible a direct inversion of the shear distribution,
simultaneously taking into account all the redshift information
(\cite{tay01}; \cite{hu02}; \cite{bac02}; paper II). Using this
technique, mass maps can also be created directly in three
dimensions.  As discussed in paper II, a mass-selected cluster
catalogue can then be extracted from these maps.  Using the SNAP
deep survey, this will result in a fine mass resolution even at
reasonable distances. Along with cosmological probes, such a
catalog can test astrophysical processes and the  hypothesis of
structure formation via gravitational instability.

\subsection{Wide Survey} \label{wide}

The SNAP mission will also include a 5 month wide survey designed
primarily for weak lensing.  This is the survey that will allow us
to use weak lensing to put constraints on cosmological parameters.
This survey will also be useful for a variety of other studies
requiring high resolution wide field multi-band imaging.

\subsubsection{Instrumental Constraints} \label{constraints}

The minimum exposure time of SNAP is constrained by the amount of
solid--state storage on the spacecraft and the ability of the
spacecraft to download data.  These two constraints have been set
at 350 GB of storage which can be downloaded once every 3 day
orbit.  This limits exposure times to  500 seconds or longer if
all filters are to be used and only lossless on-board compression
is done. Because this data set will be of great use to the larger
astronomical community, and we will utilize all nine bands to
calculate photometric redshifts for the source and lens galaxies,
we opt to collect data in all 9 bands and not to further compress
the data on-board. SNAP will be able to perform a slew and a
CCD/HgCdTe readout in about 30 seconds. Thus, our \emph{de facto}
time between exposures will be 530 seconds.

Each CCD bank consists of 9 CCDs, each with  $3510\times3510$
pixels.  Each pixel is 0.1 arcseconds square. All four CCD banks
thus provide
\[4\times9\times(3510)^{2}\times(0.1'')^2=0.34 \textrm{ deg}^{2}\] of survey
area. As with any high-orbit space mission, a high rate of cosmic
rays has been budgeted for in the SNAP orbit, and we will need to
take four dithered exposures at each pointing.  These will be
dithered by a small (a few pixels) non-integer pixel value.  A
small dither is optimal for removing cosmic rays/pixel defects;
and the non-integer pixel value allows for later ``{\tt
DRIZZLE}ing'' to increase image resolution \cite{drizzle}. To
cover each filter in the bank, we need to step the telescope six
times (either horizontally or vertically) by the size of the
optical filters. In doing so, the infrared filters are also
stepped across the field of view at the same time. Thus, the total
minimum time needed for each 0.34 square degree patch is
$4\textrm{(dithers)}\times6\textrm{(filters)}\times530\textrm{s}=12720$~seconds,
or 0.147 days.

\subsubsection{PSF Calibration}

We will need to constantly monitor the PSF through the examination
of non-saturated stars in our survey.  Calibrations with a higher
surface density of stars will need to be performed on occasion as
well.  We anticipate that we will need to perform this calibration
at least at the beginning and end of our survey, and each time
there is a focus change in the telescope. The calibration will be
done by pointing at a stellar field (such as a globular cluster,
and open cluster or a low galactic latitude field), and taking 4
dithered images for each CCD bank. There are four banks of CCDs,
requiring $4\times4\times530 \textrm{ seconds}=0.1$ days for each
full PSF calibration.  The predicted observing efficiency of the
telescope is 86\% including the time needed for downloading data
and time spent not observing while passing through radiation
zones.  If we estimate that we will need to perform one
calibration every 2 months, this requires less than 0.2\% of the
telescope time during a weak lensing survey. Thus, we estimate
that approximately 85\% of the time allotted to a weak lensing
survey will be used to gather data.

\subsubsection{Survey Characteristics}

Given 85\% efficiency, it takes 0.17 days to observe a 0.34 square
degree patch.  Therefore, we can observe 100 square degrees in 50
days.  Paper III demonstrates that, for constraining cosmological
parameters with lensing, the width of the survey is more important
than its depth.  We therefore select the minimum 500s individual
integration time at the hard limit of onboard storage and download
rate given in \S\ref{constraints}. Thus, given 5 months of time,
or 150 days observing time, our optimal survey will be:

\begin{itemize}
\item 300 square degrees \item 6 optical and 3 infrared filters
\item 2000 seconds integration in each optical filter \item 4000
seconds integration in each infrared filter \item 4 dithers to
improve resolution and eliminate cosmic rays
\end{itemize}

2000 seconds of exposure time allows us to reach a 5$\sigma$ point
source detection limit for an object with  27.5 in $I$ and 28.0 in
$V$. For a 10$\sigma$ detection of an extended galaxy with an
exponential profile, as is relevant to weak lensing, these limits
drop to isophotal magnitudes of 26.0 and 26.4 respectively.

According to paper II, this depth allows us to measure the shapes
of $\sim120$ galaxies per square arcminute in the $I$ band.
Photometric redshifts can be calculated for almost all of these
with an  error of $\Delta z\approx0.05$. Co-adding 2 or more of
the bands will allow a deeper study with a higher surface density
of galaxies. Further simulations are underway to quantify the
gains available using field co-addition.

\section{Systematic Effects} \label{systematic}

The primary goal of a weak lensing survey is to measure the shapes
of as many galaxies as possible as accurately as possible. The
size, anisotropy, and temporal stability of a telescope's PSF are
the most important factors in determining the number density of
galaxies that can be measured and the accuracy with which the
shapes can be ascertained. In order to accurately measure galaxy
shapes and sizes, it is necessary to remove the effects of
telescope PSF and detector induced shear from the galaxy images.


\subsection{Contributions to the PSF} \label{psf}

In Table~\ref{contributionstopsf} we identify 8 effects which will
contribute to the  PSF and the sizes of those effects. The sizes
and shapes of the effects are estimates using SNAP engineering
models, but will be present in any wide field space based mission.
The one dimensional rms contributions from each source are listed
in arcseconds.  For circularly symmetric patterns, the
contribution is the projection of the distribution onto the $x$ or
$y$ axis. For more complicated distributions, it is 71\% of the
root sum square of the two axes. The purpose of this section is to
discuss these effects and their time variability to determine how
they impact weak lensing measurements.

\begin{figure}
 \includegraphics[width=15cm]{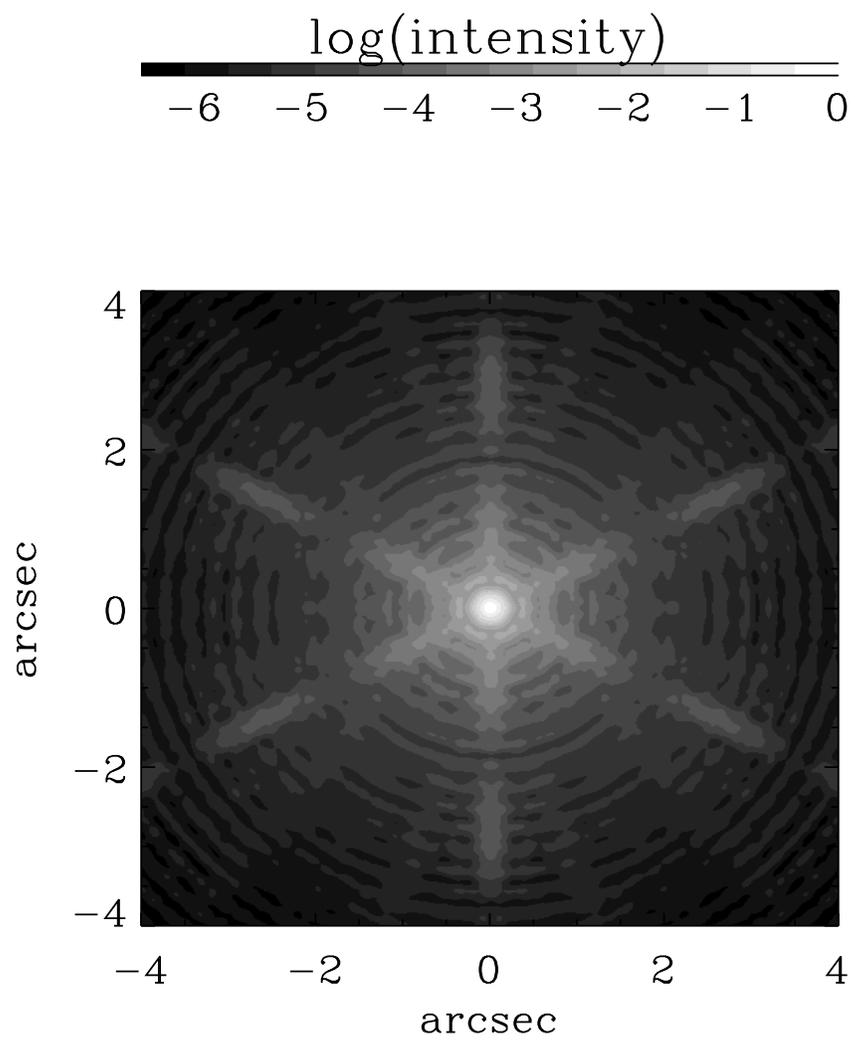}
 \caption{Oversampled image of the SNAP PSF at 800nm. The
 image is 8 arcseconds (80 SNAP pixels) on a side. Note the logarithmic
 intensity scale.}
 \label{fig:psflog}
\end{figure}

We have created model PSFs across the SNAP field of view (FOV)
taking into account some of these effects (see figure
\ref{fig:psflog}). These models will be used to study how small
perturbations of the SNAP telescope design and operating
conditions will affect the PSF. The first three items in
Table~\ref{contributionstopsf} (diffraction, diffusion, and ideal
geometric aberrations) are included in the PSF models discussed
below.  These are the most important contributions to the optical
PSF.

Item 4 (attitude control system  jitter) is difficult to model
because the sources of telescope jitter are stochastic events
caused by many different processes. The effects of jitter will
have to be measured via stellar images in orbit. Items 5 and 6
(mirror manufacturing and alignment errors) are not possible to
predict, but also can be corrected with stellar images after SNAP
has been launched. In \S\ref{misalign}, we show how slight
perturbations in mirror alignment will affect the PSF.

Item 7 (charge transfer efficiency; CTE) is a detector effect.
Electron traps within the semiconductor array are created by high
energy cosmic ray hits, and cause charge trailing during CCD
readout. This can falsely elongate all objects in the readout
direction. The magnitude of the effect can vary across the CCD,
and CTE is known to degrade over the lifetime of the mission (see
\S\ref{variability}). Tests indicate that the CTE in the Berkeley
designed CCDs being used for SNAP will be quite small and the
degradation will be significantly less than is seen on \emph{HST}
\cite{beb02a} and \cite{beb02b}.  There will not be a CTE effect
on the NIR detectors. As long as the small CTE effects are linear,
this small effect should be correctable in software using data
taken in orbit.

Item~8 (silicon transparency) is also referred to as ``red
defocus.'' This contributes only about 1 micron (0.01 arcseconds)
to PSF size at a wavelength of 800nm, and less at shorter
wavelengths . Red defocus is a consequence of the fact that blue
light is absorbed at the surface of the CCD while red light is
absorbed throughout the thickness of the CCD.  Thus, there is no
optimal focal plane for red light. This is only a problem in the
extreme red ($> 800$nm) and thus does not effect galaxy shape
measurements done using optical wavelengths.

\subsection{PSF Simulations} \label{simulations}

We have developed an {\tt IDL} routine to model the SNAP PSF
across the SNAP field of view. The PSF model takes into account
three of the effects in Table~\ref{contributionstopsf}:
diffraction from the struts and the aperture (item~1),  Gaussian
charge diffusion within the CCDs (item~2), and the spot diagram
(ray tracing through the optics; item~3). We use the currently
planned technical specifications for SNAP. The simulations are
based on a primary mirror radius of 1 meter, a secondary structure
obscuration of radius 0.4 meters (the secondary mirror itself has
radius 0.225 meters), 3 supporting struts of 4 cm thickness, and a
distance of 2.1 meters between the primary and secondary mirrors.
A  CCD diffusion value of 4.0~$\mu$m RMS is used. We use a
fiducial wavelength of 800~nm to test the effects on the PSF of
perturbations of several SNAP parameters. Below, we explore the
dependence of PSF on wavelength for optical wavelengths. We do not
explore the infrared PSF because infrared images will not be used
to measure galaxy shapes.

\begin{figure}
\includegraphics[width=15cm]{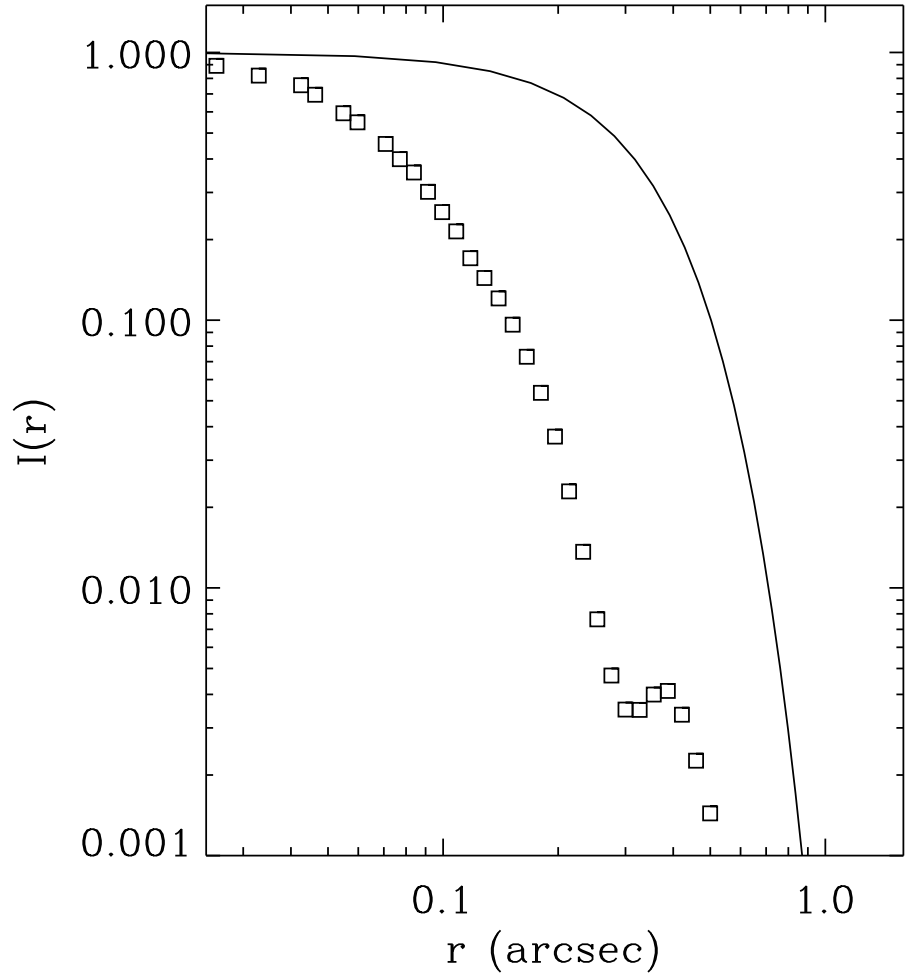}
 \caption{The averaged radial profile of the SNAP PSF at 800~nm (boxes). The curve is a
 Gaussian with FWHM 0.55 arcseconds, the best seeing consistently available
 with the Keck Telescope on the ground.}
 \label{fig:psf_profile}
\end{figure}

Figure \ref{fig:psflog} shows an oversampled PSF created a
distance of 0.01 radians ($0.57^\circ $) from the optical center
of the SNAP FOV, with an input wavelength of 800~nm. The image
measures approximately $8\times8$ arcseconds. The PSF shows a
nearly-circular central core as well as the extended diffraction
pattern caused by the struts and the aperture. Figure
\ref{fig:psf_profile} shows the average radial profile of this
PSF. The PSF intensity drops to 10\% of the central value within
0.2 arcseconds or 2 SNAP pixels. This figure also demonstrates the
improvement in PSF size of a space-based telescope over the best
ground-based PSF consistently available.

\subsection{PSF Size} \label{size}

\begin{figure}
 \includegraphics[width=15cm]{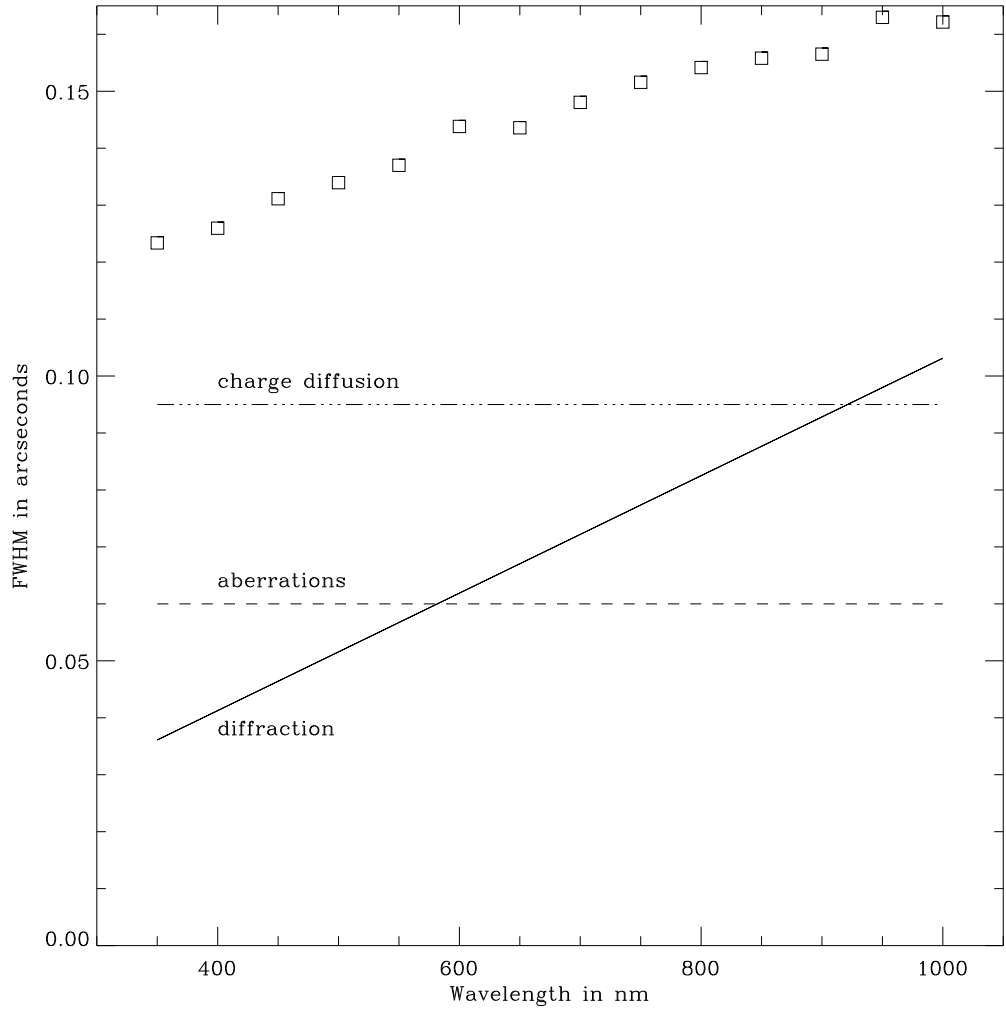}
 \caption{The FWHM of a Gaussian fit to the SNAP PSF as a function of
 wavelength (boxes). The solid line is a diffraction limited PSF
 for the 2~meter  SNAP
 primary mirror. The sizes of the effects of diffusion and aberrations are also shown.}
 \label{fig:psfsize}
\end{figure}

The size of the PSF is crucial for weak lensing because only
resolved galaxies, with sizes larger than the PSF, can provide
useful shape or size information. Figure \ref{fig:psfsize} shows
the PSF size as a function of wavelength. The size shown is the
FWHM of a Gaussian fit to the PSF by the {\tt IDL} procedure {\tt
Gauss2dfit}. Because the size of the PSF increases with increasing
wavelength, it would be advantageous for us to measure galaxy
shapes with a short wavelength. However, a higher surface density
of  galaxies can be imaged in redder filters than in bluer
filters. This is an issue that will be optimized using the
simulations described in paper II. Figure~\ref{fig:psfsize} also
shows the size of a diffraction limited PSF. Clearly, the SNAP PSF
size is not diffraction limited and is dominated by charge
diffusion and other factors.

\begin{figure}
 \includegraphics[width=15cm]{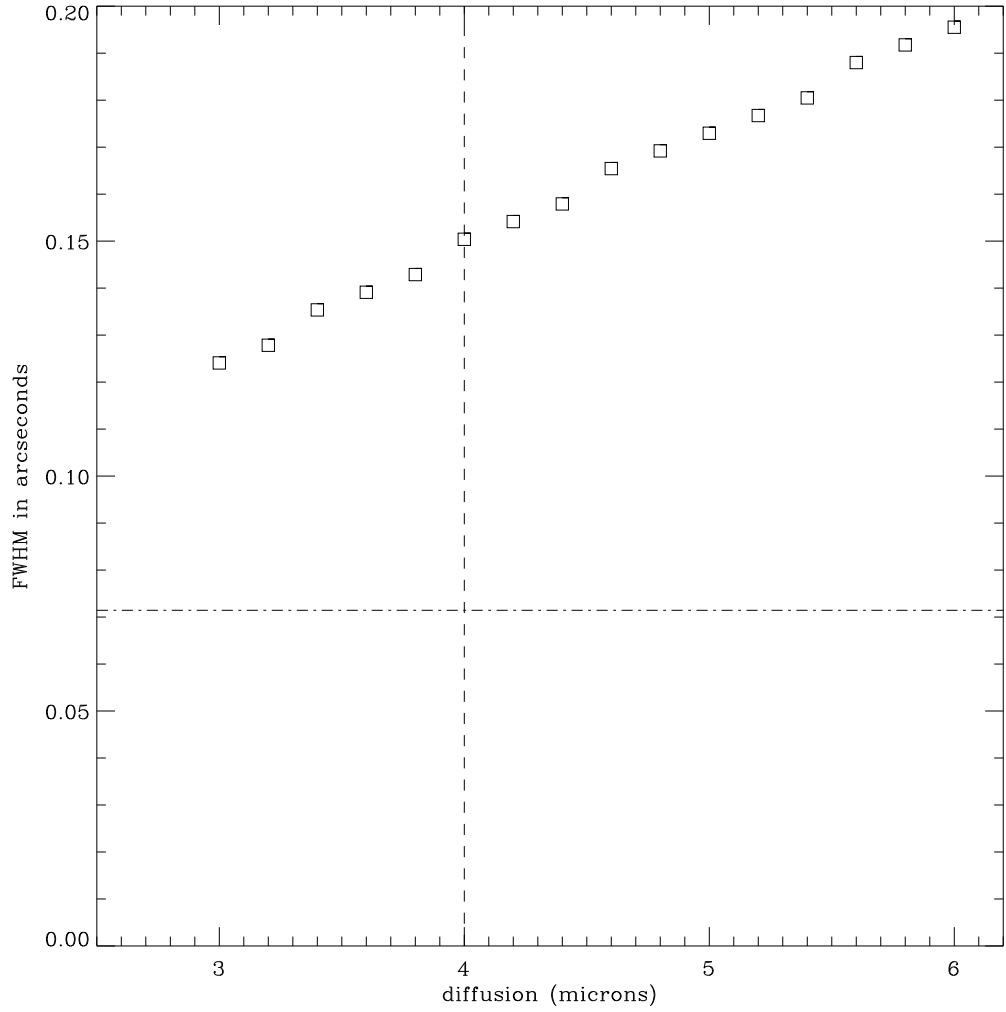}
 \caption{The FWHM of a Gaussian fit to the SNAP PSF as a function of the rms
 charge diffusion by electrons in the CCD. Higher values of diffusion are safer
 and less costly to achieve, but lead to larger PSFs. The horizontal line shows
 the FWHM of the PSF in the limit of no charge diffusion or other factors (the diffraction limit).
 The vertical line represents the default value of diffusion we use in our simulations.
 These values assume
 a wavelength of 800nm. }
 \label{fig:psfsize_diff}
\end{figure}

The rms value of charge diffusion by electrons in CCDs is driven
by the applied voltage and the thickness of the fully depleted
CCD. A higher applied voltage, or a thinner CCD, results in a
lower value of charge diffusion, benefitting the PSF. On the other
hand, a higher voltage or a thinner CCD produces a smaller
manufacturing yield, a higher failure rate, and less quantum
efficiency towards extreme red wavelengths. However, as the
allowed diffusion value increases, the size of the PSF increases
almost linearly, as shown in figure~\ref{fig:psfsize_diff}.  There
will be a detailed trade-off
study done to determine what value of diffusion strikes the proper
balance between  mission risk, cost, and weak lensing capability.
Current SNAP specifications call for a charge diffusion of
$4~\mu$m.

\subsection{PSF Anisotropy} \label{anisotropy}

The lensed shapes of galaxies that we are trying to measure are
unfortunately altered again during observation. Instrumental
effects within a telescope must be undone during data reduction in
order to recover the true image shapes and the lensing-induced
ellipticity (or polarization). The two main detector effects are
{\it smear}, or PSF convolution, and {\it shear}, which includes
astrometric distortions.

Smearing tends to limit the size of the smallest galaxies able to
be measured: size and shape information for galaxies smaller than
the PSF is lost during convolution. The isotropic component of the
PSF circularizes galaxies, while an anisotropic component may also
cause galaxies to become preferentially elongated in one
direction. Another important factor affecting the measured image
shapes is distortion from the detector. Such astrometric
distortions precisely mimic shear by weak gravitational lensing.
Detailed analysis of the SNAP detectors' geometric distortion
awaits more advanced detector models and ground measurements of
the detectors themselves.   Fortunately, the small shear
distortions predicted for SNAP should be straightforward to
subtract, using measurements of the astrometric shifts of dithered
stellar images. Furthermore, detector distortion affects only the
shape of the measured objects, rather than the size. Thus, this
effect is not a limiting factor in the size of galaxies which can
be measured.

Several techniques have been developed to correct image shapes for
both of smear and shear using software, including {\tt KSB}
\cite{kai95}, {\tt RRG} \cite{rho00} and ``shapelets''
\cite{ref03a} \cite{ref03b}; see also a related method by
\cite{ber02}. To measure the ellipticity of the PSF we
 first calculate the intensity weighted second moments $I_{xx}$,
$I_{yy}$ and $I_{xy}$. These are defined as the following sum over
pixels (\emph{i})

\begin{equation}
 I_{xy}=\frac{\sum_{i}I(x_i,y_i)x_i y_i w(x_{i},y_{i})}
             {\sum_{i}I(x_i,y_i)}
\end{equation}

\noindent where $I(x,y)$ is the intensity in a pixel, $x$ and $y$
are the distances from that pixel to the centroid of the PSF and
$w(x,y)$ is a Gaussian weighting function with a standard
deviation of 0.2 arcseconds (two SNAP pixels). Similar equations
hold for $I_{xx}$ and $I_{yy}$. Following lensing convention, the
two-component ellipticity $e_{i}$ is defined as

\begin{equation}
 e_1=\frac{I_{xx}-I_{yy}}{I_{xx}+I_{yy}} \hspace{1cm}
 e_2=\frac{2I_{xy}}{I_{xx}+I_{yy}}.
\end{equation}

\noindent Measured ellipticities can then be corrected for
instrumental distortion using higher order weighted moments, and
the moments of the PSF (see  \emph{e.g}.~\cite{rho00}).

\begin{figure}
\includegraphics[width=15cm]{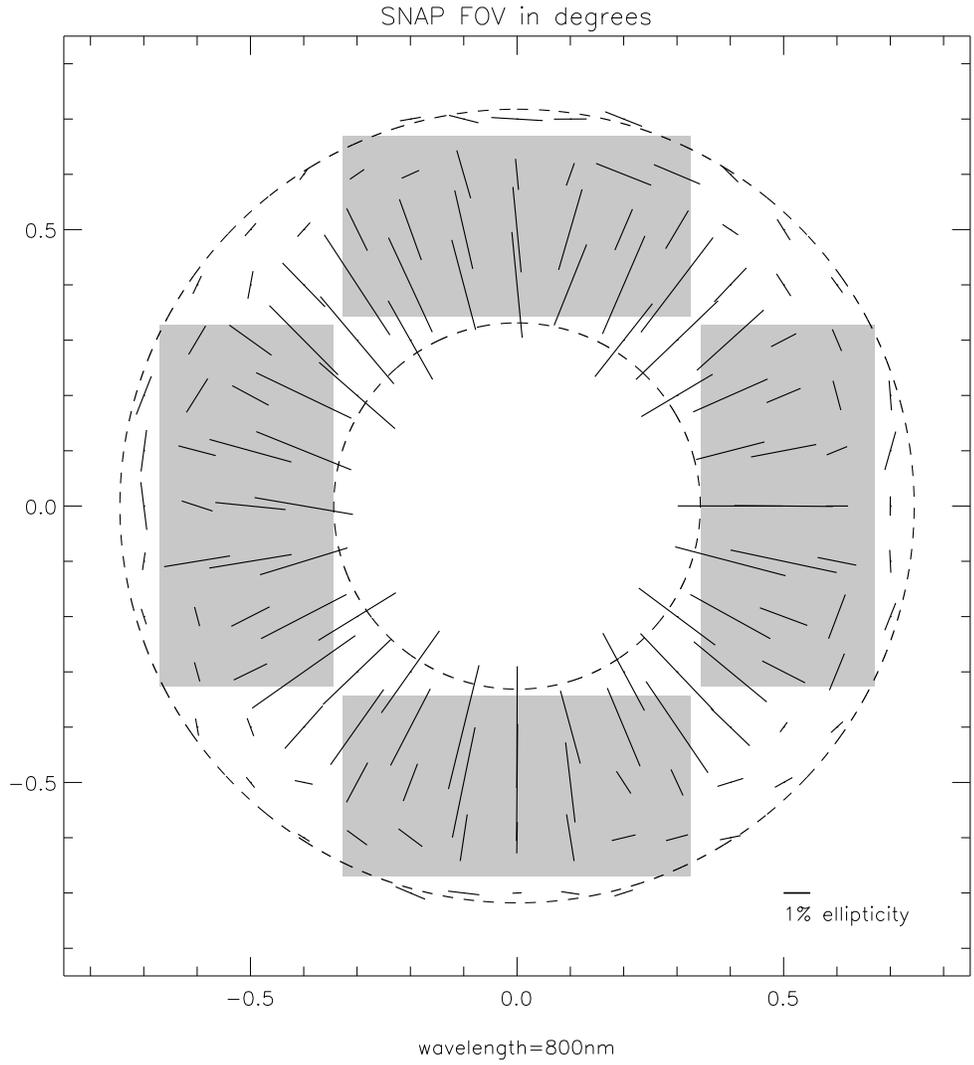}
 \caption{The PSF induced ellipticity over the SNAP FOV at 800 nm. Each line represents the size of the
 ellipticity that the PSF induces in a point--like source at that
 position.  This
 ellipticity field is wavelength dependent so the actual measured PSF would
 depend on the fixed filter at a given position.}
 \label{fig:psf_ellip80}
\end{figure}

Figure \ref{fig:psf_ellip80} shows the ellipticity of the SNAP PSF
over the SNAP field of view produced by the first three factors
listed in Table~1 at a wavelength of 800~nm. The size of the PSF
induced ellipticity is not large, roughly 4--5 \% at most. For
comparison, the PSF induced ellipticity of WFPC2 on HST is up to
10\%, as measured empirically \cite{rho00} and calculated using
the program \texttt{TINYTIM} \cite{kri03}. The PSF does change
over the focal plane and in fact over a single CCD detector.
Therefore, a much finer grid of model PSFs would be needed to
accurately model the SNAP PSF over the entire focal plane.

\subsection{Mirror Misalignment} \label{misalign}

The above PSF simulations were performed assuming a perfect mirror
alignment. The effects of a simple mirror misalignment can be
added to the simulations by creating a new spot diagram for the
misaligned mirrors. Such a misalignment may occur due to thermal
fluctuations in the barrel of the telescope, and particularly in
the secondary mirror support struts.  SNAP engineering estimates
indicate that the mirror alignment error will be at maximum
$\theta=2\times10^{-4}$ degrees.  Most likely the mirror
misalignment would be only half of that.

\begin{figure}
 \includegraphics[width=15cm]{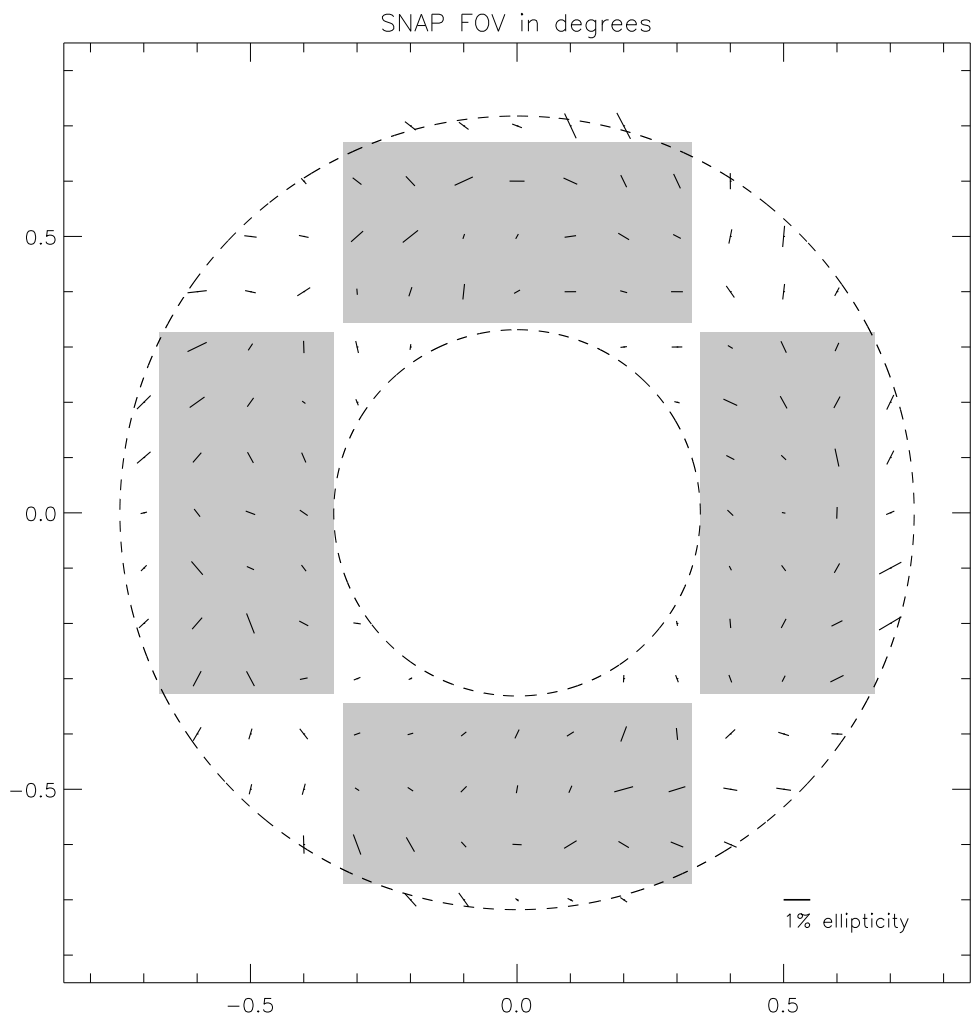}
 \caption{The change in the PSF induced ellipticity between the ideal mirror
 alignment and a situation in which the secondary mirror alignment is tilted by
 $\theta=10^{-4}$ degrees. }
 \label{fig:diff_0_1}
\end{figure}

\begin{figure}
\begin{center}
 \includegraphics[width=15cm]{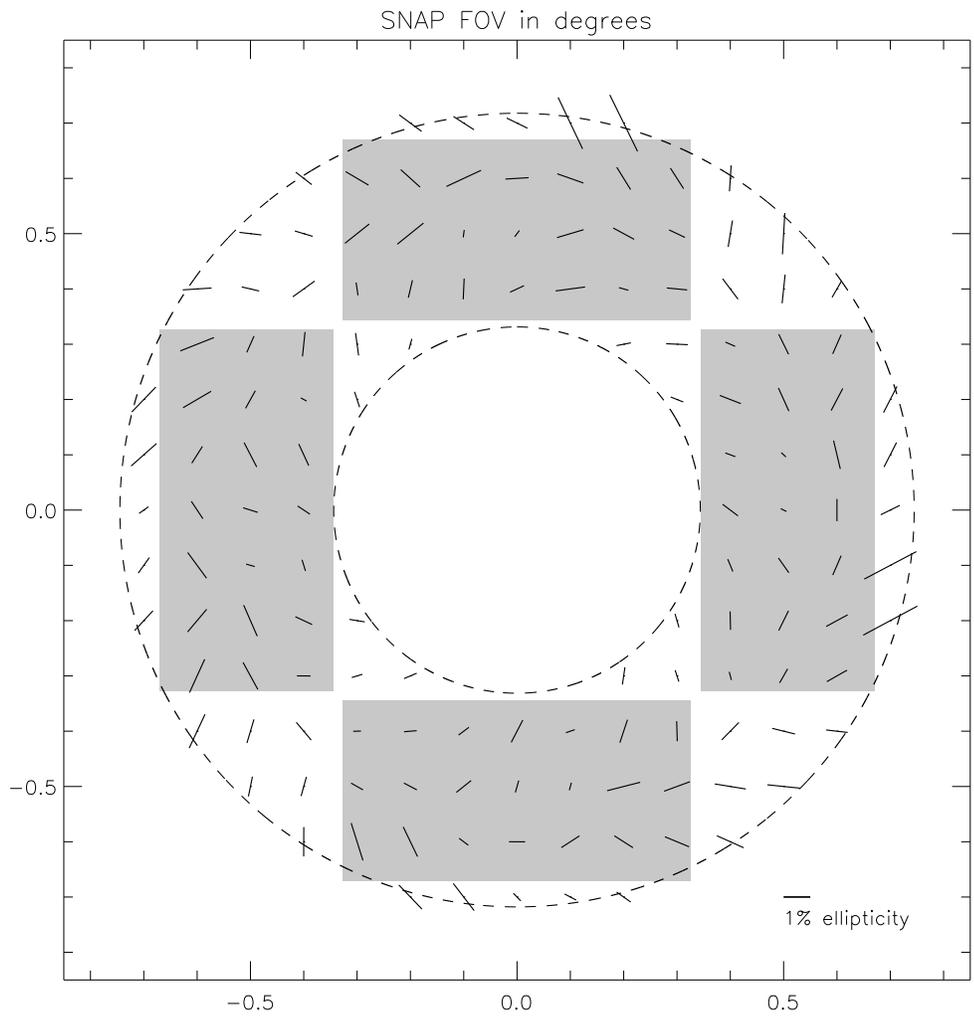}
\end{center}
 \caption{Same as figure~\ref{fig:diff_0_1} but with a mirror tilt of
 $\theta=2\times10^{-4}$ degrees. }
 \label{fig:diff_0_2}
\end{figure}

Figures~\ref{fig:diff_0_1} and \ref{fig:diff_0_2} show the change
in the induced ellipticity caused by mirror alignment errors of 1
and $2\times10^{-4}$ degrees, respectively. These plots indicate
how much the induced ellipticity would differ from the nominal
perfectly aligned mirrors in figure~\ref{fig:psf_ellip80}, and as
such are shown for a wavelength of 800nm. These plots represent
the maximum error SNAP would face if the mirrors become misaligned
and no correction is made to galaxy shapes for the misalignment.
This error manifests itself as a residual post-correction rms
ellipticity $\langle(\Delta e)^{2}\rangle^{\frac{1}{2}}$ where
$\Delta e$ is the difference in ellipticity
$e=\sqrt{(e_1^{2}+e_2^{2})}$ of the PSF between the aligned and
misaligned mirrors, and the angle brackets indicate an average
over the SNAP FOV. This residual ellipticity is 0.5\% for a mirror
alignment error of $10^{-4}$ degrees and 0.9\% for an alignment
error of $2\times10^{-4}$ degrees.  For comparison, the typical
residual ground-based post-correction ellipticity is 5-10\%. Thus,
in the worst-case scenario when a mirror alignment error goes
unnoticed, this effect will only introduce an error five to ten
times smaller than that found in  ground-based images. Vigilant
monitoring of the SNAP PSF will allow us to correct for mirror
misalignment and reduce this error.

\subsection{Other Sources of Time Variability} \label{variability}

The time variability of the PSF is a concern because of the
accuracy to which object shapes need to be measured for weak
lensing.  The HST's PSF changes significantly in time periods of
order days and even changes during the course of its ninety minute
orbit \cite{hoe98} \cite{rho00}, hindering corrections for
instrumental shape distortions. In addition to possible mirror
misalignment errors discussed above, SNAP will suffer from some
amount of ``structural dryout creep'' which is an outgassing of
water from carbon fiber elements of the optical support structure.
The optical supports will shrink as this outgassing occurs, but
this is expected to last only a few months and then stabilize.
During this initial phase, the telescope will be refocused to
bring the PSF back to its nominal value but the PSF will drift
away from that value as the telescope goes out of focus.

There will also be an initial thermal contraction for several
months, and possible ``creaking'' of the detector support
structure, as the telescope cools after launch. Thus, this will
not be the optimal time for weak lensing measurements. Throughout
its lifetime, the spacecraft will also undergo further thermal
contraction and expansion cycles as the solar exposure changes
during its orbit. The currently planned highly elliptical orbit
will minimize this effect, but the consequences upon the PSF will
have to be monitored by examining stellar data.

The charge transfer efficiency (CTE) of CCDs is known to degrade
over time, as cosmic ray hits create electron traps within the
semiconductor array. These traps will cause image trailing during
CCD readout, falsely elongating all the galaxies in the readout
direction. This is clearly a concern for weak lensing.
With this in
mind, the SNAP CCDs are being specifically designed to undergo
minimal CTE degradation.


\subsection{PSF Based Survey Requirements} \label{requirements}

Based on the above analysis of the SNAP PSF, we plan the following
for a weak lensing survey:

\begin{enumerate}
\item The wide field weak lensing survey should not be conducted
within
      the first several months of  launch
\item Galaxy shapes should be measured with a filter at $\sim
800$~nm or shorter to
      utilize the smaller PSF
\item Stellar images (in low Galactic latitude fields) should be
taken at regular intervals to monitor and correct for
      the  PSF
\item Astrometric shifts of stars should be used to calculate
detector distortion
      early in the mission
\item Aim for 4.0~$\mu$m diffusion or less as a trade-off between
mission
      risk, cost, and PSF size
\end{enumerate}

\section{Conclusions} \label{conclusions}

A wide field space telescope is  crucial in the drive to
understand both dark energy and dark matter. We have studied the
systematic effects contributing to the PSF of such a telescope.
The PSF can be designed to be much smaller than the best available
from the ground, and more stable over time than ground-based PSFs
or even that of the HST. These high quality image specifications
ensure that a telescope like SNAP will be a powerful instrument
for the next generation of precision weak lensing experiments. We
have outlined baseline survey strategies that will lead to
exciting new lensing results.

Paper II introduces simulations of space-based images that are
being used to predict the sensitivity to weak gravitational
lensing, using the specifications presented here. That paper
includes the accuracy and resolution of possible dark matter maps.
Paper II also contains a calculation of the accuracy of
photometric redshifts in SNAP data. These numbers are then applied
in paper III to determine how well wide field space based
observations will be able to constrain cosmological parameters
including the dark energy equation of state parameter {\it w}.

 JR was supported by an NRC/GSFC Research Associateship. AR was
supported in Cambridge by an EEC fellowship from the TMR network
on Gravitational Lensing, by a Wolfson College Research
Fellowship, and by a PPARC advanced fellowship. We thank the
Raymond and Beverly Sackler Fund for travel support.  We thank the
anonymous referee and Douglas Clowe for useful suggestions.

\clearpage

\begin{table*}
 \begin{center}
  \caption{Model Contributions to the SNAP PSF.\label{contributionstopsf}}
  \vspace{1mm}
  \nopagebreak
  \begin{tabular}{|p{0.1cm}|p{6cm}|p{7cm}|}
   \hline
   \# & Effect & Size of PSF Contribution \\
  \hline\hline
   1 & optical diffraction & circular Airy disk 0.06 arcsec RMS at 1000~nm \\
   2 & electron diffusion & circular Gaussian 0.04--0.05 arcsec RMS \\
   3 & ideal geometric aberrations & blobs 0.02--0.03 arcsec RMS \\
 4 & attitude control system jitter & circular Gaussian 0.02 arcsec RMS\\
   5 & mirror manufacturing errors & circular Gaussian 0.02 arcsec RMS \\
   6 & mirror alignment errors &  circular Gaussian 0.02 arcsec RMS\\
   7 & charge transfer efficiency & linear $< 0.01$ arcsec\\
   8 & transparency of silicon (red defocus) &  linear $< 0.01$ arcsec\\
  \hline
  \end{tabular}
 \end{center}
\end{table*}
\normalsize



\end{document}